\documentclass{aa}  

\usepackage{graphicx}
\usepackage{subcaption}
\usepackage[dvipsnames]{xcolor}
\usepackage{soul}
\newcommand{\firstcite}[1]{\citeauthor{#1} (\citeyear{#1}; hereafter referred to as \citetalias{#1})}
\defcitealias{Orosz2011}{O+11}
\defcitealias{Coriat2012}{C+12}
\defcitealias{Fragos2015}{F+15}
\usepackage{txfonts}
\begin{document}

   \title{Understanding the evolution of black hole spin in X-ray binary systems: Case study of XTE~J1550-564}

   \titlerunning{A case study of the XTE~J1550-564 system's black hole spin}

   \author{ L. Bartolomeo Koninckx\thanks{Fellow of the Consejo Nacional de Investigaciones Cient\'ificas y T\'ecnicas (CONICET), Argentina, E-mail: leandrobart96@fcaglp.unlp.edu.ar}
          \and
            M. A. De Vito\thanks{Member of the Carrera del Investigador Cient\'ifico, CONICET, Argentina.}
          \and
            O. G. Benvenuto\thanks{Member of the Carrera del Investigador Cient\'ifico, Comisi\'on de Investigaciones Cient\'ificas de la Provincia de Buenos Aires (CIC), Argentina.}
          }

   \institute{
   Instituto de Astrof\'isica de La Plata, IALP, CCT-CONICET-UNLP, Argentina and Facultad de Ciencias\\ 
   Astron\'omicas y Geof\'isicas de La Plata, Paseo del Bosque S/N, (1900) La Plata, Argentina.\\
}

    \authorrunning{L. Bartolomeo Koninckx,  M. A. De Vito, and O. G. Benvenuto} 

   \date{Received March 28, 2025; accepted September 9, 2025}
   
 \abstract
   {We present a comprehensive study of the X-ray binary system XTE~J1550-564, with the primary objective of analyzing the evolution of the black hole's spin parameter. To achieve this objective, we embarked on the necessary step of identifying a plausible progenitor for the system. Using a set of models covering various parameter combinations, we were able to replicate the system's observed characteristics within acceptable error margins, including fundamental parameters such as component masses, orbital period, donor luminosity, and effective temperature. The model results indicate the possibility of diverse evolutionary pathways for the system, highlighting the significant role played by the initial mass of the donor star and the efficiency of mass transfer episodes. While some models are well-aligned with estimates of the mass transfer rate, they all fall short of explaining the black hole's observed moderate spin ($a^* = 0.49$). We also explored alternative magnetic braking prescriptions, finding that only an extreme and fully conservative scenario, based on the convection and rotation boosted prescription, can reproduce the observed spin and only in a marginal way. Our study attempts to shed light on the complex dynamics of black hole X-ray binaries and the challenges of explaining their observed properties with theoretical models.}

   \keywords{stars: evolution -- stars: low-mass -- stars: black holes -- X-rays: binaries -- stars: individual: V381Normae --  X-rays: individuals: XTE~J1550-564}

   \maketitle
%

\section{Introduction}\label{Sec:intro}
In binary systems, one particularly interesting classification is the ``close binary,'' where the orbital separation between its components facilitates recurrent episodes of mass transfer. If the accretor happens to be a compact object, this system is designated as an X-ray binary. These kinds of systems have captured substantial scientific attention ever since their initial discovery due to their distinctive and noteworthy characteristics \citep{Cowley92,Verbunt93,KRETSCHMAR2019,Bahramian2022}.\\

 Close binary systems reveal an intriguing composition, made up of a mass-losing (donor) star and a black hole (BH). When discussing rotating BHs, the parameters used to comprehensively characterize them are their mass,  $M_{\mathrm{BH}}$,
and spin angular momentum, $J_{\mathrm{BH}}$. For such systems, we can occasionally measure both of these parameters at once. The first,  ,  $M_{\mathrm{BH}}$,
can be obtained from a study of the orbital motion of the donor star with an independent measurement of the mass of the star and the inclination of the system. Measuring the spin angular momentum is more challenging, as it requires the study of relativistic effects in the region near the compact object. Once the mass of the compact object is ascertained, the BH spin angular momentum can be delineated through the dimensionless parameter, $a^* = a/M_{\mathrm{BH}} = cJ_{\mathrm{BH}} / GM_{\mathrm{BH}}^2$, where $G$ represents the universal gravitational constant and $c$ is the vacuum speed of light. Observationally deducing this parameter relies on two main methodologies: the continuum-fitting approach \citep{Zhang97} and the relativistic reflection method \citep{Fabian89, Laor91}. The former involves modeling the thermal X-ray spectrum originating from the accretion disk, while the latter is focused on modeling the profile of the K$\alpha$ iron line. It is worth noting that when these two methods are applied to the same object, the resulting estimations do not necessarily overlap \citep{2013MNRAS.431..405R}. There are just three known systems where the results are, in fact, compatible, namely, GRS 1915+105 \citep{McClintock2006,Blum2009}, Cyg X-1 \citep{Gou2011}, and XTE~J1550-564 \citep{Steiner2011}.

As the BH accretes matter, its dimensionless spin parameter increases. Therefore, computing the entire mass transfer episode using theoretical models provides a way to understand the evolution of this fundamental parameter. In a previous paper, we presented a detailed analysis of the binary V404~Cyg to identify a plausible progenitor of the system and to calculate the evolution of the spin parameter of the accreting BH \citep{Bartolomeo2023}. For this purpose, we assumed that the BH was  not rotating initially, finding that our evolutionary models could not otherwise achieve the high estimate of its spin parameter, $a^*>0.92$ \citep{Walton2017}.

This result has motivated a new study, where we aim to determine the evolution of the BH spin parameter for another system, XTE~J1550-564. This system has a higher confidence level in measuring the BH spin parameter and a much lower value, which should be easier to achieve with the assumption of an initially non-rotating BH. Considering this fact and understanding its fundamental characteristics makes the system XTE~J1550-564 an excellent candidate for exploring whether it is possible to measure the BH spin through accretion alone, assuming the BH was not rotating initially. To accomplish this goal, we need to construct a robust evolutionary model for the system, from which the mass transfer episodes can be studied and, thus, the evolution of the BH spin parameter can be determined.\\

A general study addressing the origin of BH spin in LMXBs was performed by \firstcite{Fragos2015}. They analyzed nine BH X-ray binaries (BHXBs), including XTE J1550-564, for which the dimensionless BH spin parameter was measured, to study their evolution through accretion. As a standard assumption,  the BH was  not rotating initially and on this basis, they found that the dimensionless BH spin values of these systems could be explained. However, they only considered  the case of fully conservative mass transfer models; in other words, they focused only on the scenario
where the BH accretes all the mass lost by the donor star. Additionally, the mass accretion rate was not Eddington-limited. They considered this limit and lower values for the accretion efficiency in the post-processing of the data in an approximate way, but this approach is not self-consistent with the orbital evolution, which is still assumed to follow the initial models. On the other hand, in \citetalias{Fragos2015} only the dimensionless BH spin parameter measurements via continuum fitting were used, which for XTE J1550-564 indicated that a lower value  needed to be achieved for this parameter.\\
\cite{Reynolds2021} stated that assuming conservative mass transfer is an important caveat in \citetalias{Fragos2015}. Many of the BHXBs included in that work are observed to possess winds that are originated in the inner accretion disk \citep{Miller2015}. Moreover, evidence of an ionized disk wind in the XTE~J1550-564 system has been found by \cite{2020ApJ...892...47C}. 
The evolution of the BH spin relies strongly on the amount of mass that the BH is accreting and also on the orbital evolution. On the other hand, if possible, the value being aimed for should be a confident value measured with both techniques. 
This motivated us to also explore the non-conservative mass transfer scenario in this study, where the models are Eddington-limited and computed with a self-consistent orbital evolution, while an observational estimation for the BH spin parameter is consistently measured by both techniques.\\
The structure of this work is as follows. We summarize the main characteristics of the XTE~J1550-564 system in \S~\ref{sec:XTE}. We provide insights into the numerical code in \S~\ref{sec:code}, followed by a description of the methods employed in \S~\ref{sec:methods}. We present the models and their outcomes in \S~\ref{sec:results}. Subsequently, we discuss the astrophysical implications arising from the obtained results in \S~\ref{sec:discussion} and present our concluding thoughts in \S~\ref{sec:conclusions}.

\section{XTE J1550--564}\label{sec:XTE}
XTE~J1550-564 was first identified on September 7, 1998, over the course of observations made with the All-Sky Monitor (ASM) on board the \textit{Rossi X-Ray Timing Explorer} satellite \citep{Smith98}. During this period, the system experienced a significant outburst, resulting in a substantial increase in its X-ray emissions, making it detectable by X-ray observatories. The optical counterpart was later identified as V381~Normae at the coordinates R.A. $= 15^{\mathrm{h}} 51^{\mathrm{m}} 04^{\mathrm{s}}$ and Decl. $= -56^{\circ} 28' 37.5"$ \citep{Orosz98}. Subsequently, through spectroscopic observations, \citet{Orosz2002} conclusively confirmed the compact object's identity as a BH, contributing to the growing catalog of confirmed BHXBs discovered during that period. 
In a study conducted by \firstcite{Orosz2011}, a dynamical model was developed using spectroscopic and photometric data obtained from the 6.5~m Magellan Telescopes at Las Campanas Observatory. The authors successfully derived fundamental parameters of the system, including an orbital period of $P_{\mathrm{orb}} = 1.5420333$~days, an inclination angle of $74^\circ.7 \pm 3^\circ.8$, and component masses $M_\mathrm{d} = 0.3 \pm 0.07 \,\mathrm{M}_\odot$ and $M_{\mathrm{BH}} = 9.1 \pm 0.61 \,\mathrm{M}_\odot$ for the donor star and the BH, respectively. In addition to these findings, the authors provided an estimation for the donor's luminosity, denoted as $L_{\mathrm{d}}$, yielding a value of $1.05^{+0.6}_{-0.33}\,\mathrm{L}_\odot$, alongside an effective temperature, $T_{\mathrm{eff}}$, corresponding to a K3III spectral type of 4475~K. The fundamental parameters used for this study are summarized in Table~\ref{tab:Obs}.

\begin{table}
    \caption{Values adopted for the fundamental parameters describing the system XTE~J1550-564.}
    \centering
    \renewcommand{\arraystretch}{1.3}
    \begin{tabular}{cccc}
        \hline\hline
        \textbf{Parameter} & \textbf{Value} & \textbf{Error} & \textbf{Unit}\\\hline
        M$_{\mathrm{d}}$ & 0.3 & $\pm0.07$ & M$_\odot$\\
        M$_{\mathrm{BH}}$ & 9.1 & $\pm0.61$ & M$_\odot$\\
        P$_{\mathrm{orb}}$ & 1.5420333 & $\pm0.0000024$ & d \\
        L$_{\mathrm{d}}$ & 1.05 & $^{+0.6}_{-0.33}$ & L$_\odot$\\
        T$_{\mathrm{eff}}$& 4475 & $\pm250$ & K\\
        \hline\hline        
    \end{tabular}
    \label{tab:Obs}
\end{table}

Theoretical investigations have been performed to analyze the mass transfer rate within the donor star of XTE~J1550-564. \citetalias{Orosz2011} employed Equation~(33) from the work of \citet{King88}, in conjunction with the core mass derived from their self-developed dynamical model. Through this methodology, they derived an approximate mean mass loss rate of $\langle\dot{M}_{\mathrm{d}}\rangle \approx 9 \times 10^{-11}$ M$_\odot\, \mathrm{yr}^{-1}$. Conversely, \firstcite{Coriat2012} assumed parity between the average accretion rate onto the BH and the mass loss rate from the donor star. By utilizing the average X-ray luminosity obtained from observations, they estimated the aforementioned parameter as $\langle\dot{M}_{\mathrm{d}}\rangle \approx 1.6 \times 10^{-9}$ M$_\odot \, \mathrm{yr}^{-1}$. It is important to note that this estimation was obtained using values for the BH mass and system distance that differ from those reported by \citetalias{Orosz2011}. These two parameters are crucial for the calculations conducted in their study. It is also noteworthy that these determinations differ from each other by more than an order of magnitude.

As already mentioned, there are two techniques for measuring the dimensionless BH spin parameter, $a^*$. In the case of XTE~J1550-564, both methods were employed by \citet{Steiner2011} and yielded congruent results. The derived value of $a^* = 0.49^{+0.13}_{-0.20}$ indicates a moderate spin, implying that the relativistic jets within this system are predominantly fueled by the accretion disk, rather than the rotational spin of the BH \citep{Steiner2011}.
\section{Binary evolution code}\label{sec:code}
In this study, we present theoretical simulations conducted using our binary evolution code, which has been comprehensively detailed in prior works \citep{Benvenuto2003,DeVito2012,BDVHa}. Our calculations encompass the entire evolutionary timeline of the donor star, with the foundational assumption that the BH formation occurs before the donor star attains the zero-age main sequence (ZAMS). The simulations extended up to an age of 14 Gyr. In conjunction with evolutionary calculations, our code performed a detailed analysis of the orbital dynamics governing the system. We operated under the premise that both components follow circular trajectories. The components remained detached unless either of them approached the Roche lobe, as defined within the framework of the circular restricted three-body problem. During phases of detachment, when the component radius remained below the approximation of the Roche lobe radius based on \citet{Eggleton83}, our code worked as a conventional one. However, when this detachment condition was not met, signifying a mass transfer episode, the code incorporated the mass transfer rate as an additional variable in the differential equations. We assumed  that these episodes might not conserve mass and angular momentum within the system. Instead, they were parameterized using the prescriptions of \citet{Rappaport82,Rappaport83}, relying on two free parameters: 1) the fraction $\beta$ of mass lost by the donor star that the companion accretes and 2) the specific angular momentum of matter expelled from the system, denoted as $\alpha$ and measured in units relative to the same quantity for the compact object. In addition to mass loss, we also considered that the system can lose angular momentum via gravitational radiation \citep{Landau71} and magnetic braking (MB).

For the latter, we considered three different prescriptions. Our first models were computed employing the Skumanich Law (hereafter, MB0, \citealt{1972ApJ...171..565S}), which is one the most commonly used for binary evolution calculations \citep{Verbunt81,Rappaport83}. In this case, the time derivative of the angular momentum due to MB0 can be written as
\begin{equation}\label{eq:MB0}
    \dot{J}_{\mathrm{MB0}} = -3.8\times10^{-30}M_1 R_\odot^4 \left(\frac{R_1}{R_\odot}\right)^4\omega_1 \mathrm{~dyn~cm,}
\end{equation} where $M_1$, $R_1$, and $\omega_1$ are the mass, radius, and angular velocity of the donor star\footnote{We designate the donor star with the subscript~1, and the BH with the subscript~2. It is important to note that we will use these numerical subscripts to represent the quantities obtained as outputs of our models.}. Hereafter, the subscript $\odot$ represents a defined quantity, this time for the Sun. We also considered the MB3 prescription \citep{2019MNRAS.483.5595V}, where the time derivative of the angular momentum loss due to MB3 also considers the convective turnover timescale $(\tau_\mathrm{conv})$ and the donor's mass loss rate by stellar winds $(\dot{M}_{\mathrm{1,wind}})$ as 
\begin{equation}\label{eq:MB3}
        \dot{J}_{\mathrm{MB3}} = \dot{J}_{\mathrm{MB0}} \left( \frac{\tau_\mathrm{conv}}{\tau_{\odot,\mathrm{conv}}} \right)^{2}\left( \frac{\dot{M}_{1,wind}}{\dot{M}_{\odot,wind}} \right).
\end{equation}Finally, we explored the Convection and Rotation Boosted MB (CARB, presented in \citealt{2019ApJ...886L..31V}) that also takes into consideration the effect of the stellar rotation in the Alfvén radius, expressed as
\begin{align}\label{eq:CARB}
    \dot{J}_{\mathrm{CARB}} = -\frac{2}{3} \dot{M}_{\mathrm{1,wind}}^{-{1}/{3}} R_1^{{14}/{3}} \left( v_{\mathrm{esc}}^2 + \frac{2\omega_1^2 R_1^2}{K^2}  \right)^{-{2}/{3}} \nonumber\\ \times~\omega_\odot B_\odot^{{8}/{3}}\left( \frac{\omega_1}{\omega_\odot} \right)^{{11}/{3}} \left( \frac{\tau_\mathrm{conv}}{\tau_{\odot.\mathrm{conv}}} \right)^{{8}/{3}}.
\end{align}Here, $v_\mathrm{esc}$ is the surface escape velocity of the donor, $B_\odot$ represents the magnetic field magnitude of the Sun ($B_\odot = 1$~G), and $K=0.07$ is a constant obtained from a grid of simulations made by \citet{2015ApJ...798..116R}.
The reasons for considering alternative MB prescriptions will be addressed in \S \ref{sec:discussion}.

For the calculations presented in this study, we maintained a constant value of $\alpha=1$, while we explored different values for $\beta$ as part of the parameter space. In terms of elemental abundances, we adopted the solar metallicity for the donor star in the system, with the specific composition characterized by X~=~0.721, Y~=~0.265, and Z~=~0.014. In addition, the mixing length parameter was  fixed at a value of $\alpha_{\mathrm{MLT}} = 1.51$. These values were determined to provide the best representation by calibrating the actual Sun with our code. As for the alternative MB prescriptions, in this work, we took the values of $\omega_\odot = 3\times10^{-6}$~s$^{-1}$, $\tau_{\odot, \mathrm{conv}} = 1.537\times10^6$~s$^2$, and $\dot{M}_{\odot,\mathrm{wind}}=2.54\times10^{-14}$~M$_\odot$~yr$^{-1}$ \citep{2006ima..book.....C}. We refer to \citet{Bartolomeo2023} for a more comprehensive elucidation of our code and the underlying assumptions guiding this study and to \citet{Echeveste2024} for the specifications of our code regarding alternative MB prescriptions.
\section{Methods}\label{sec:methods}
To identify a plausible progenitor for the XTE~J1550-564 system using our code, we initiated an extensive exploration of the parameter space. Specifically, we varied the initial parameters, which encompassed the initial masses of both the donor star and the BH, the orbital period, and the value of $\beta$.

Initially, we delineated a broad parameter range, subsequently refining it into a more detailed grid of values. In this specific investigation, our exploration encompassed the following parameter values: for the initial masses of the donor star, we considered 1.1, 1.25, and 1.4~M$_\odot$; for the initial masses of the BH, we employed a range from 8.4 to 9.1~M$_\odot$ in increments of 0.1~M$_\odot$; the $\beta$ parameter was assessed at values of 0.1, 0.5, and 1.0; in other words, we  chose to let the system be strongly non-conservative, moderately non-conservative, or fully conservative. In this set of calculations, we considered the MB0 prescription. We also explored other MB prescriptions in \S~\ref{MBs}. The initial orbital period was contingent upon the initial mass of the donor star. Given the system's proximity to the bifurcation orbital period value\footnote{We remark that this is the case for the physical ingredients we have assumed here. If we consider another MB prescription (see, e.g., \citealt{Echeveste2024}), this may no longer be the case.}, this parameter significantly influenced the modeling outcomes. Consequently, we considered orbital periods ranging from 1.14 to 1.20 days for an initial donor star mass of 1.1~M$_\odot$, from 0.80 to 0.85 days for 1.25~M$_\odot$, and from 0.64 to 0.70 days for 1.4~M$_\odot$. In all instances, the increment used was 0.01 days. With all these considerations, the total number of models created amounted to 480.\\


After computing all the models within the grid, our binary evolution code provides the time evolution of key parameters for each model. To identify which models best represent the observed system, we performed a $\chi^2$ test at each time, 
\begin{equation}\label{eq:epsilon}
\chi^2 = \sum^{5}_{i=1} \chi^2_i,\hspace{0.3cm}
\chi_i = \frac{Q_{i,obs} - Q_{i,mod}}{\sigma_{i,obs}},
\end{equation} where $Q_i$ represents the parameters of interest (orbital period, component masses, donor luminosity, and effective temperature). Here, $Q_{i,obs}$ and $\sigma_{i,obs}$ denote the observed values and their respective uncertainties (see Table \ref{tab:Obs}), while $Q_{i,mod}$ corresponds to the modeled values at a given age. The analysis was conducted with five degrees of freedom and a significance level of $0.05$, allowing us to distinguish models that closely match observations from those that did not. A model was considered to be representative of the XTE~J1550-564 system if all parameters simultaneously fell within their observational error margins. By applying this criterion and ranking models based on their reduced $\chi^2$ values, we identified the ones that best fit the system's observations.

Once the progenitor was identified, our model provides insights into the accretion history of the BH. As the compact object accretes matter, it undergoes a process of spin-up. Therefore, this knowledge is a crucial component in computing the evolution of the BH's spin parameter. To estimate this evolution, we adopted  formula (10) from the work of \citet{Podsiadlowski2003}, operating under the assumption that the BH was born with no initial rotation,
\begin{equation}\label{BHspinparameter}
    a^*(t)=\left( \frac{2}{3} \right)^{\frac{1}{2}} \frac{M^0_{\rm BH}}{M_2(t)} \left\{ 4-\left[ 18\left( \frac{M^0_{\rm BH}}{M_2(t)} \right)^2 -2\right]^{\frac{1}{2}} \right\}.
\end{equation}This equation holds true for cases where $M_2 < \sqrt{6}M^0_{\rm BH}$ \citep{Bardeen70,King99}, a condition satisfied throughout our calculations. In this equation, $M^0_{\rm BH}$ represents the initial mass of the BH, while ${M_{\rm 2}(t)}$ corresponds to the modeled mass at the specific time, $t$, during the evolutionary process.
\section{Results}\label{sec:results}
\begin{table*}
    \centering
    \caption{Models falling within all observational uncertainties.}
    \begin{tabular}{c|cccc|ccc}
         \hline\hline
         \textbf{Model} & $\mathbf{M^0_{d}} [\mathrm{M}_\odot]$ & $\mathbf{M^0_{BH}} [\mathrm{M}_\odot]$ & $\mathbf{P^0_{orb}} [\mathrm{d}]$ & $\mathbf{\beta}$ & $\mathbf{t_{\mathrm{obs}}} [\mathrm{Myr}]$ & $\mathbf{\chi^{2}(t_{\mathrm{obs}})}$ & $\left| \chi^2_{red} - 1 \right|$  \\\hline
         A & 1.4  & 9.0  & 0.65  & 0.5  & 6352  & 1.34 & 0.73 \\
         B & 1.25  & 8.5  & 0.80  & 0.5  & 8027 & 1.04 & 0.79 \\
         C & 1.25 & 9.1  & 0.82  & 0.5  & 7115  & 1.03 & 0.79 \\
         D & 1.1  & 8.6  & 1.16  & 0.5  & 9991  & 0.98 & 0.80 \\
         E & 1.1  & 8.8  & 1.16  & 1.0  & 10053 & 0.94 & 0.81 \\
         F & 1.4  & 8.9  & 0.65  & 0.5  & 6295  & 0.39 & 0.92 \\
         \hline\hline
    \end{tabular}
    \tablefoot{From left to right, the columns include the model name, initial masses for the donor star and BH, initial orbital period, $\beta$ parameter value, the age at which the system best falls within the observational errors, the $\chi^2$ statistic value at that moment, and the distance from the unit of the reduced chi-square ($\chi^2 / \nu$, where $\nu$ are the degrees of freedom).}
    \label{tab:models}
\end{table*}
\subsection{Modeling a progenitor for XTE~J1550-564}\label{sec:modeling}

In this section, we present our analysis for the model results within the grid, as specified in the previous section. Among these models, we have identified six evolutionary scenarios that satisfy the condition that the orbital period, component masses, donor's luminosity, and effective temperature all fall within their respective observational error ranges simultaneously.

\begin{figure}
    \centering
    \includegraphics[width = 0.8\columnwidth]{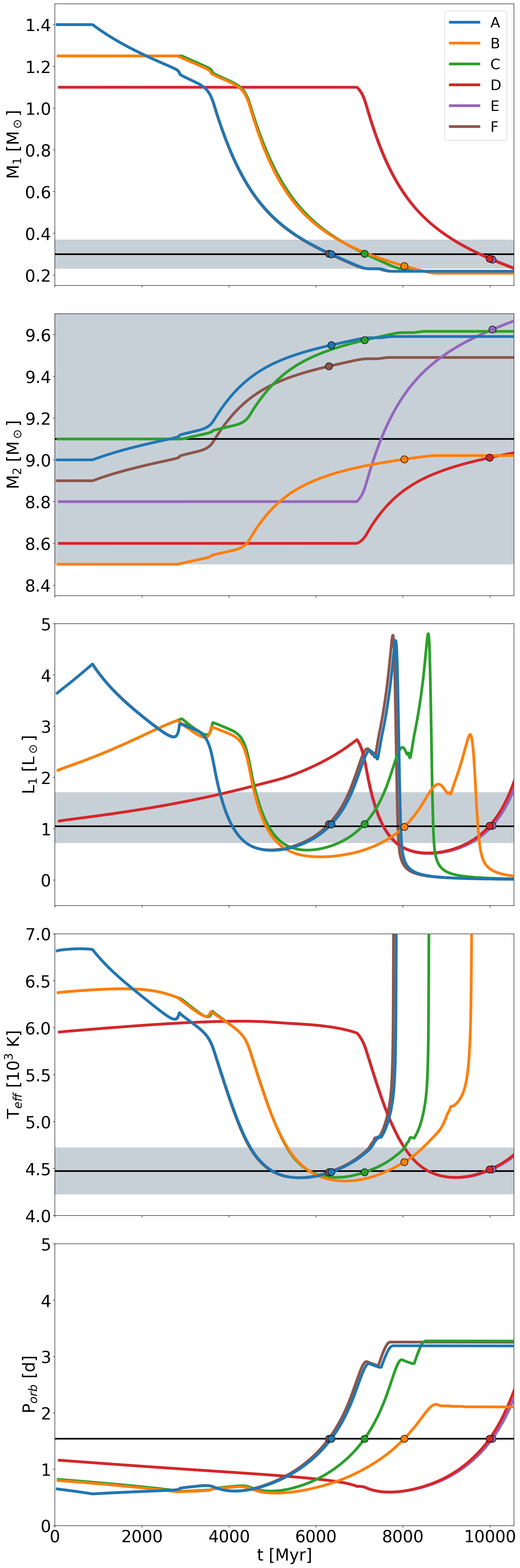}
    \caption{Evolution of fundamental quantities as a function of time for the models listed in Table~\ref{tab:models}. Observed values are indicated by horizontal black lines, with their respective errors represented by the shaded gray area. Each point on the graph corresponds to when the model reaches the ${t_\mathrm{obs}}$ value.}
    \label{fig:models}
\end{figure}

These models are presented in Table \ref{tab:models}, where we provide detailed information about the initial parameters considered for each of them, along with the specific time (${t_\mathrm{obs}}$) when all the quantities were aligned with their respective error bars. To facilitate referencing, we  assigned names to these models using letters from ``A'' to ``F'' and arranged them in ascending order of the distance to the unit of the reduced chi-squared value for ${t_\mathrm{obs}}$. The evolution of fundamental quantities for these models is presented in Figure~\ref{fig:models}. This graphical representation provides compelling evidence supporting the notion that these models can indeed be regarded as plausible progenitors for the XTE~J1550-564 system.

\begin{figure}
    \centering
    \includegraphics[width = \columnwidth]{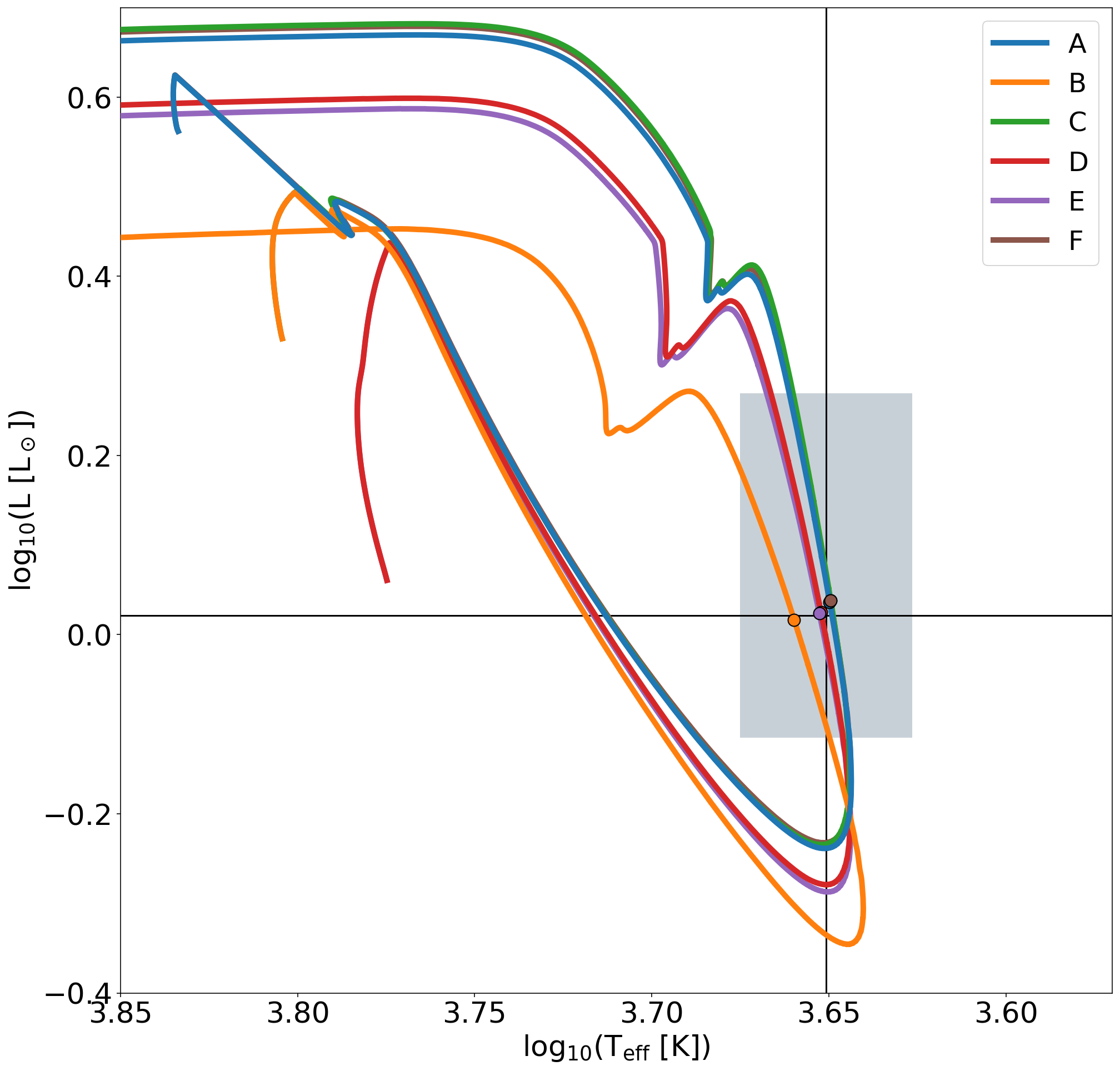}
    \caption{Evolutionary tracks of the donor star for the models representing the system XTE~J1550-564. The points on the graph represent the estimation of the actual age for each model. Black lines show the observational estimations for the donor's luminosity and effective temperature, while the respective errors are represented by the grey-shaded area.}
    \label{fig:HR}
\end{figure}

We present a portion of the evolutionary tracks of the models on the Hertzsprung-Russell diagram (HRD) in Figure \ref{fig:HR}. It is noteworthy that in all cases, the mass transfer episode commences during the hydrogen-burning phase in the core, corresponding to Case A of the mass transfer episode \citep{1967ZA.....65..251K}. At the present age, the system is progressing through the red giant branch phase, while continuing to lose mass. The ultimate stage of the donor star's evolution is anticipated to be a helium white dwarf with an approximate mass of 0.21~M$_\odot$.

\subsection{Comparison with prior mass transfer rate estimations}

Considering the significant disparity between the two available estimations for the mass loss rate, namely, that of \citetalias{Orosz2011}, $\langle\dot{M}_{\mathrm{d}}\rangle~\approx~9\times10^{-11}$~M$_\odot~\mathrm{yr}^{-1}$, and that of \citetalias{Coriat2012}, $\langle\dot{M}_{\mathrm{d}}\rangle~\approx~1.6~\times~10^{-9}$~M$_\odot~\mathrm{yr}^{-1}$, we made an independent estimation. Employing Equation (25a) of \citet{Webbink83}, we obtained a value of $\langle\dot{M}_{\mathrm{d}}\rangle \approx 1.17 \times 10^{-10}$~M$_\odot~\mathrm{yr}^{-1}$, which is very close to the one provided by \citetalias{Orosz2011}.

\begin{figure}
    \centering
    \includegraphics[width = 0.7\columnwidth]{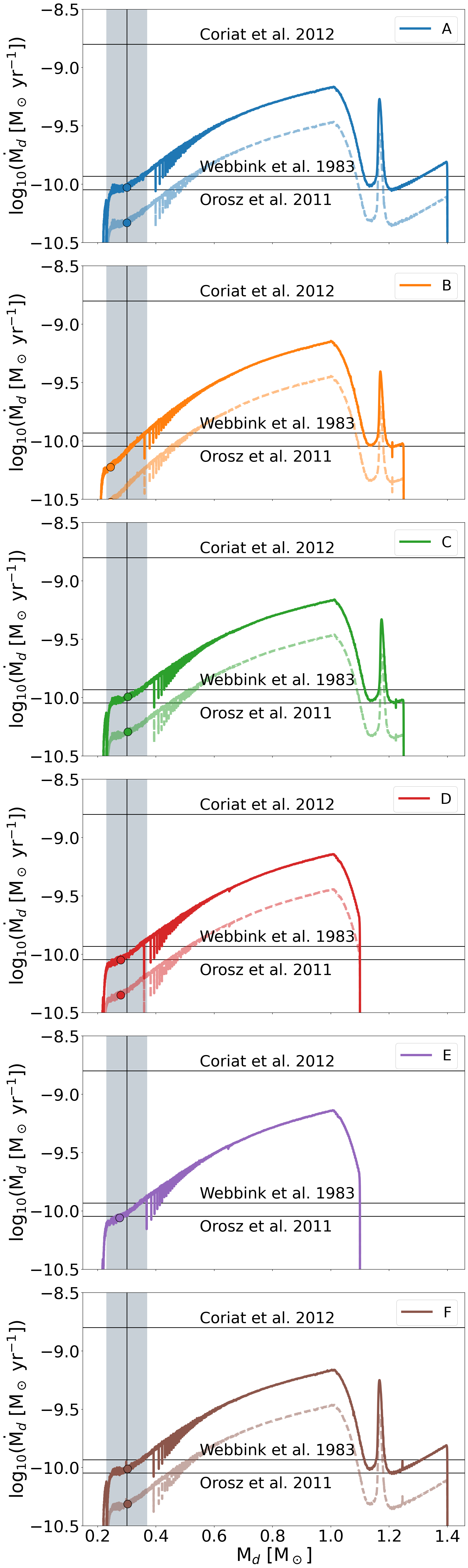}
    \caption{Absolute value of the mass loss rate (full line) and mass accretion rate (dashed line) as a function of the donor's mass for the best models. Horizontal black lines represent the estimations for the mass loss rate derived from the works of \citetalias{Coriat2012}, \citetalias{Orosz2011} and the estimation we obtained with \citet{Webbink83}. The donor's mass ($M_\mathrm{d} = 0.3$~M$_\odot$) is indicated by a vertical black line, with a shaded area representing the observational error. The dots depict the donor's mass and mass loss rate (or mass accretion rate) at $t_{\mathrm{obs}}$ for each model.}
    \label{fig:mdot}
\end{figure}

The mass transfer episodes were studied for each of the best models. In all cases, the mass transfer rates were far lower than the Eddington critical rate. The results obtained from our models for the evolution of this parameter are presented in Figure \ref{fig:mdot}, illustrating the evolution of the mass loss rate and the mass accretion rate as a function of the donor's mass. At the estimation for the present age for each model, t$_\mathrm{obs}$, the values of $\dot{M}_\mathrm{1}$ are in closer agreement with the estimates made by \citetalias{Orosz2011} and the present study, as opposed to those by \citetalias{Coriat2012}. We delve into the potential reasons for this discrepancy in \S~\ref{sec:discussion}.

Although the work of \citetalias{Fragos2015} does not specifically provide the mass loss rate at the current age of the system, Table 5 of that work details the time that has passed between the beginning of mass transfer up to the system's current age, along with the amount of mass accreted during this time for each model. This allows us to estimate an average mass transfer rate  between 2.1~$\times~10^{-10}$ and 6.9~$\times~10^{-10}$~M$_\odot$~yr$^{-1}$ for \citetalias{Fragos2015} candidate  progenitors of XTE~J1550-564. As for our models, the average mass transfer rates can also be calculated in the same way, obtaining values between 2.0~$\times~10^{-10}$ and 2.7~$\times~10^{-10}$~M$_\odot$~yr$^{-1}$.\\

\subsection{BH spin parameter evolution}
\begin{figure}
    \centering
    \includegraphics[width = \columnwidth]{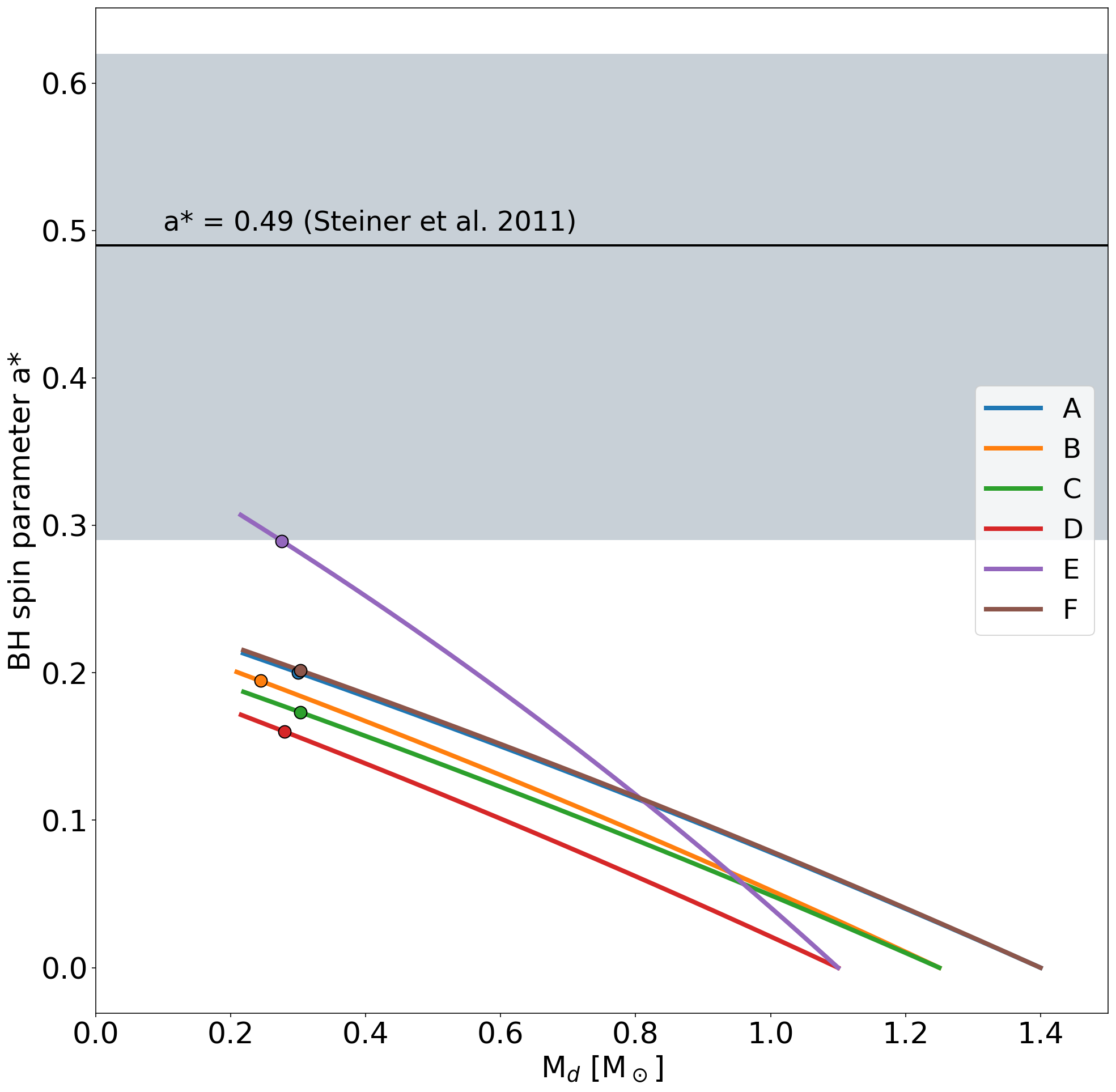}
    \caption{Evolution of the dimensionless BH spin parameter $a^*$ as a function of the donor mass. The observational data obtained by \citet{Steiner2011} of $a^* = 0.49^{+0.13}_{-0.20}$ is depicted as a black horizontal line, with the shaded gray area representing the error margin.}
    \label{fig:spin}
\end{figure}
 Having determined the detailed evolution of the mass transfer episodes for each of our models, we were able to investigate the evolution of the dimensionless BH spin parameter, $a^*$, by employing Equation (\ref{BHspinparameter}). The results for this parameter are illustrated in Figure \ref{fig:spin}. Notably, when $t_{\mathrm{obs}}$ is reached, the estimations for the BH spin parameter consistently fall well below the observational reference. The only conservative model of the ones presented here gets the highest value, reaching $a^* = 0.28$, which is right below the lower error margin of the measured spin.\\

These estimations appear to cluster into two distinct branches. This is not a mere coincidence but rather a consequence of the two values of the $\beta$ parameter that our best models have: 0.5 and 1. This parameter fundamentally governs the amount of matter available for accretion onto the BH. As more matter accretes, the BH's spin increases. Consequently, the results for the BH spin parameter are grouped according to these values of $\beta$. In a secondary order of influence, the evolution of $a^*$ also demonstrates dependence on the initial mass of the donor star. Given that the star's evolution primarily drives the occurrence of mass transfer episodes and the donor star's mass dictates the available matter for the BH to accrete, it follows that models with varying initial donor star masses but the same $\beta$ value exhibit varying estimations for the BH spin parameter. Specifically, models with more massive donor stars tend to yield higher estimations for the BH spin parameter, as shown in the figure.
\section{Discussion}\label{sec:discussion}
Our results, akin to the scope of this study, were structured around two primary objectives. The first entailed the determination of a plausible progenitor for the XTE~J1550-564 system, which we consider a fundamental step toward achieving the second objective: the analysis of the evolution of the BH's spin and its comparison with the observationally obtained value. In this section, we delve into the astrophysical insights derived from our endeavors in pursuit of these dual objectives.\\
\subsection{A progenitor for XTE~J1550-564}
In the preceding section, we present the results of our models, each of which replicated the main observational characteristics of the XTE~J1550-564 system within its respective error margins. We would naturally seek to find which of these models best represents the system. Based on the results depicted in Figure \ref{fig:models}, it becomes evident that all of the models exhibit a nice agreement with the five fundamental parameters we chose to consider. While they were initially constructed with minor input parameter variations, these models yield notably different outcomes. This is caused by the value of the orbital period being very close to the bifurcation value that separates the systems into convergent and divergent regimes.

In the subsequent phase of our analysis, we turned our attention to the mass transfer episodes. As previously discussed in \S~\ref{sec:results}, the expected mass transfer rate for XTE~J1550-564, based on our stellar evolution calculations and the estimations made by \citetalias{Orosz2011}, is an order of magnitude lower than those deduced from accretion disk arguments by \citetalias{Coriat2012}. This incongruity prompts an intriguing question regarding whether our models can sustain mass transfer rates as high as those postulated by \citetalias{Coriat2012}. This scenario might indeed be feasible if the irradiation feedback mechanism is at play or if a stronger MB prescription is considered. We discuss these possible scenarios below.

\subsubsection{Irradiation of the donor star}

Irradiation feedback occurs when, as a consequence of mass transfer from the donor star to a compact companion, it receives X-ray irradiation \citep{1993A&A...277...81H,2004A&A...423..281B,BDVHa} in return. In cases where the donor star possesses a deep outer convective zone (as is the case for XTE~J1550-564, given its relatively cool outer layers), the incoming irradiation onto the X-ray-illuminated portion of its photosphere partially inhibits the region's ability to efficiently release the energy emanating from the stellar interior. Consequently, what would otherwise be a gradual and continuous mass transfer process occurring on long timescales spanning timescales up to a few gigayears becomes unstable, manifesting itself as a sequence of episodes interspersed with prolonged periods of halted mass transfer. These episodes are short-lived, typically spanning megayears, yet the peak transfer rate is far higher than in standard calculations (see, e.g., \citealt{BDVHa}). Notably, with this mechanism, it becomes feasible to account for mass transfer rates as high as those proposed by \citetalias{Coriat2012} for XTE~J1550-564 and, potentially, even higher.

For the occurrence of pulsed mass transfer, the X-ray source must be able to provide a continuous emission over an amount of time exceeding the Kelvin-Helmholtz (thermal) timescale of the donor star, denoted as $\tau_{\rm KH}$. This condition may be satisfied if the compact companion is a neutron star, with irradiation emissions emanating from its surface. On the contrary, in the case of XTE~J1550-564, the compact object is a BH. Thus, the surrounding accretion disk is the only potential X-ray source capable of enabling feedback to occur. Remarkably, as discussed in \citet{2008NewAR..51..869R}, for the conditions inherent to the object under study here, the disk is unable to sustain emission over timescales as long as $\tau_{\rm KH}$. Consequently, we conclude that this mechanism should not be in operation in XTE~J1550-564 and, thus, irradiation feedback does not provide a plausible explanation for the very high mass transfer rate proposed by \citetalias{Coriat2012}.

\subsubsection{Other magnetic braking prescriptions}\label{MBs}
\begin{figure*}[h]
       \centering
       \begin{subfigure}{\columnwidth}
           \centering
           \includegraphics[width=\textwidth]{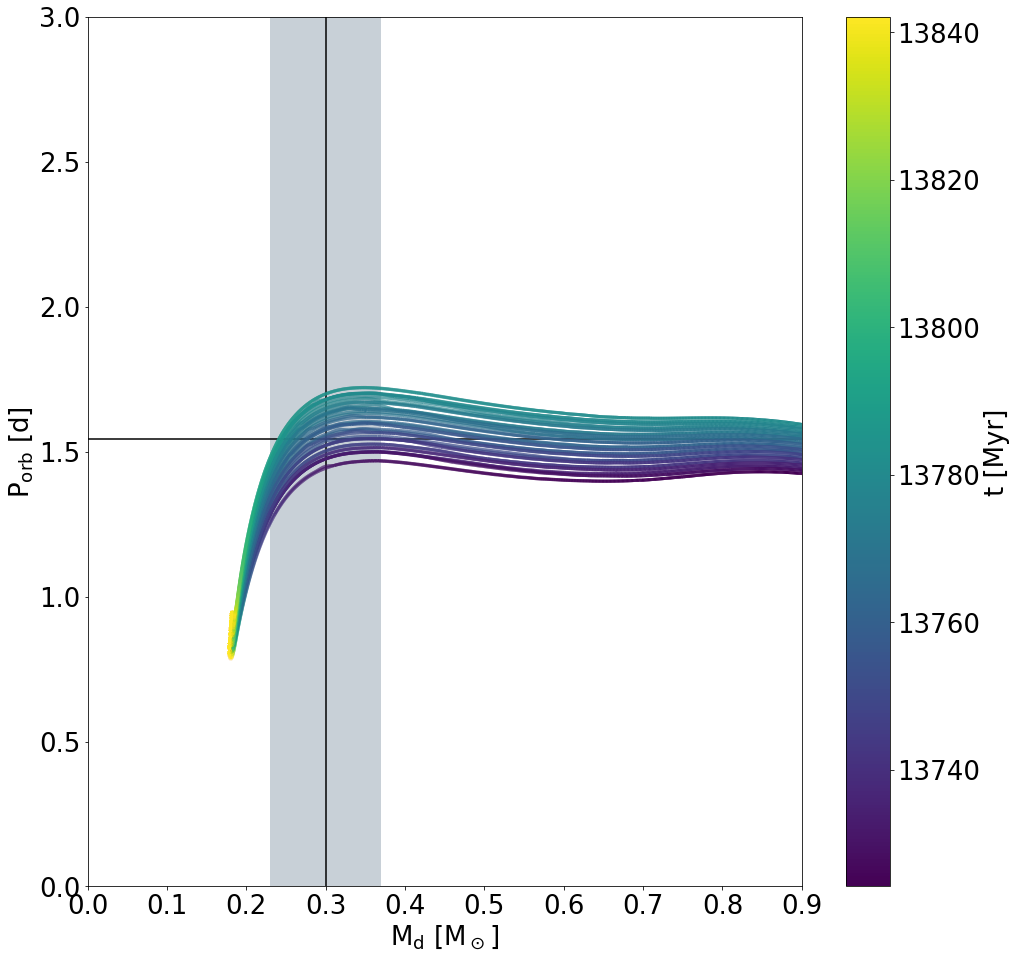}
       \end{subfigure}
       \hfill
       \begin{subfigure}{\columnwidth}
           \centering
           \includegraphics[width=\textwidth]{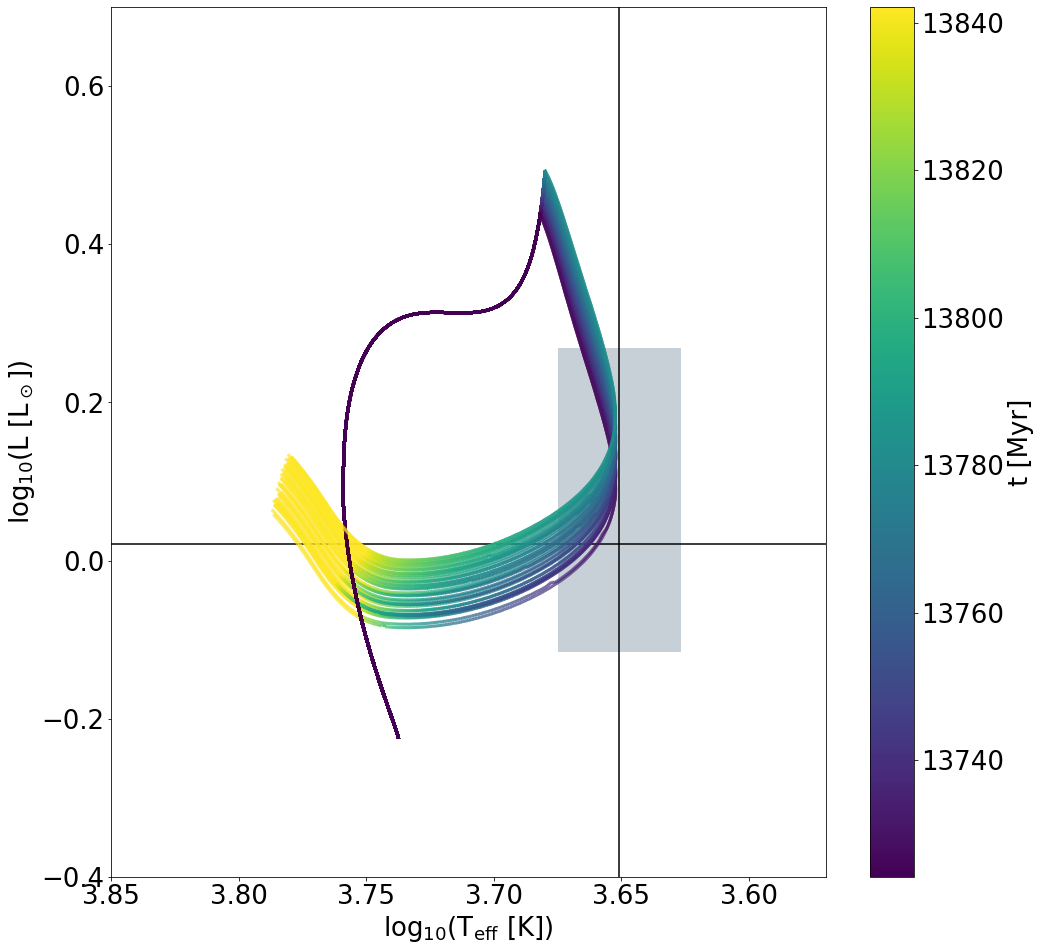}
       \end{subfigure}
       \newline
       \begin{subfigure}{\columnwidth}
           \centering
           \includegraphics[width=\textwidth]{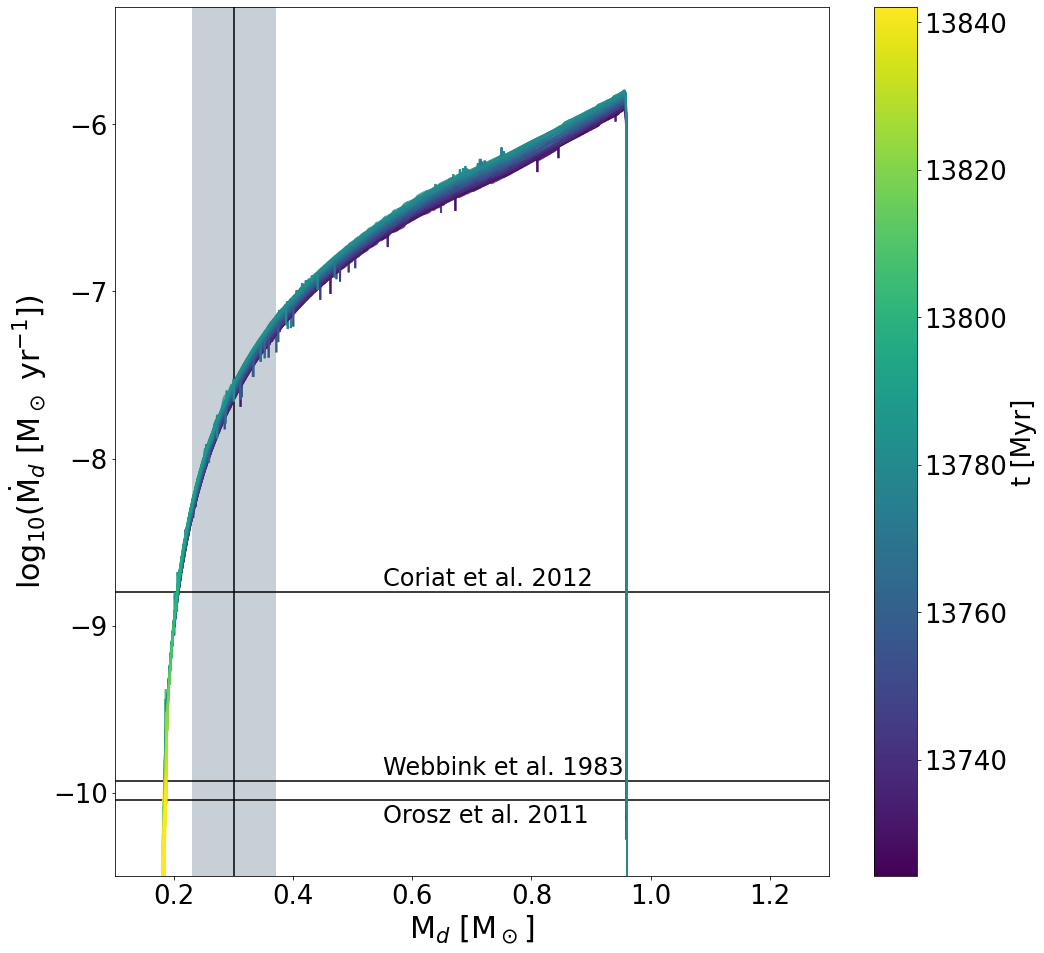}
       \end{subfigure}
       \hfill
       \begin{subfigure}{\columnwidth}
           \centering
           \includegraphics[width=\textwidth]{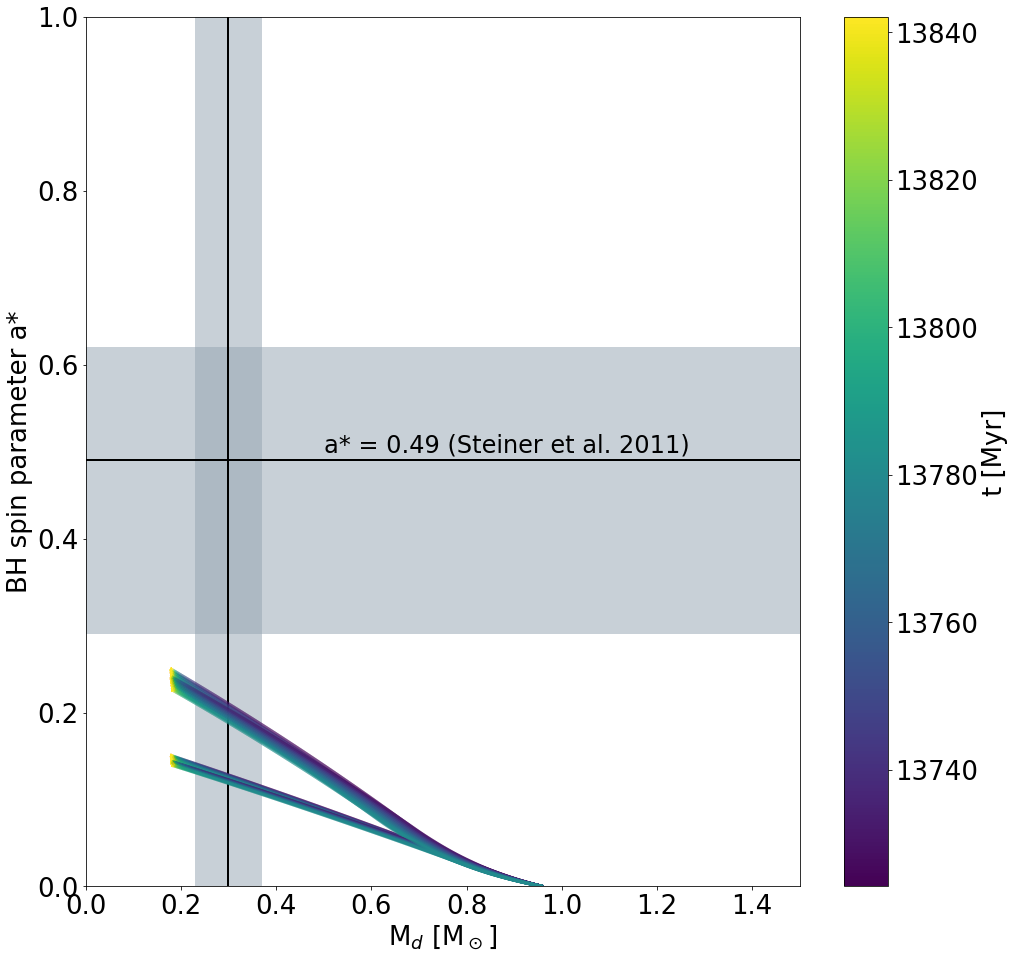}
       \end{subfigure}
       \caption{Evolution of the models computed using the MB3 prescription. \textit{Top-left}: Donor's mass vs. orbital period. \textit{Top-right}: Hertzprung-Russel diagram. \textit{Bottom left}: Donor's mass vs. mass transfer rate. \textit{Bottom right}: Donor's mass vs.  BH spin parameter. Black lines indicate the observed values of the donor mass, luminosity, effective temperature, orbital period, mass transfer rate, and BH spin parameter; shaded grey areas represent their associated uncertainties. The color gradient traces the system’s age during the mass transfer episode.}
       \label{fig:MB3}
\end{figure*}
\begin{figure*}[h]
       \centering
       \begin{subfigure}{\columnwidth}
           \centering
           \includegraphics[width=\textwidth]{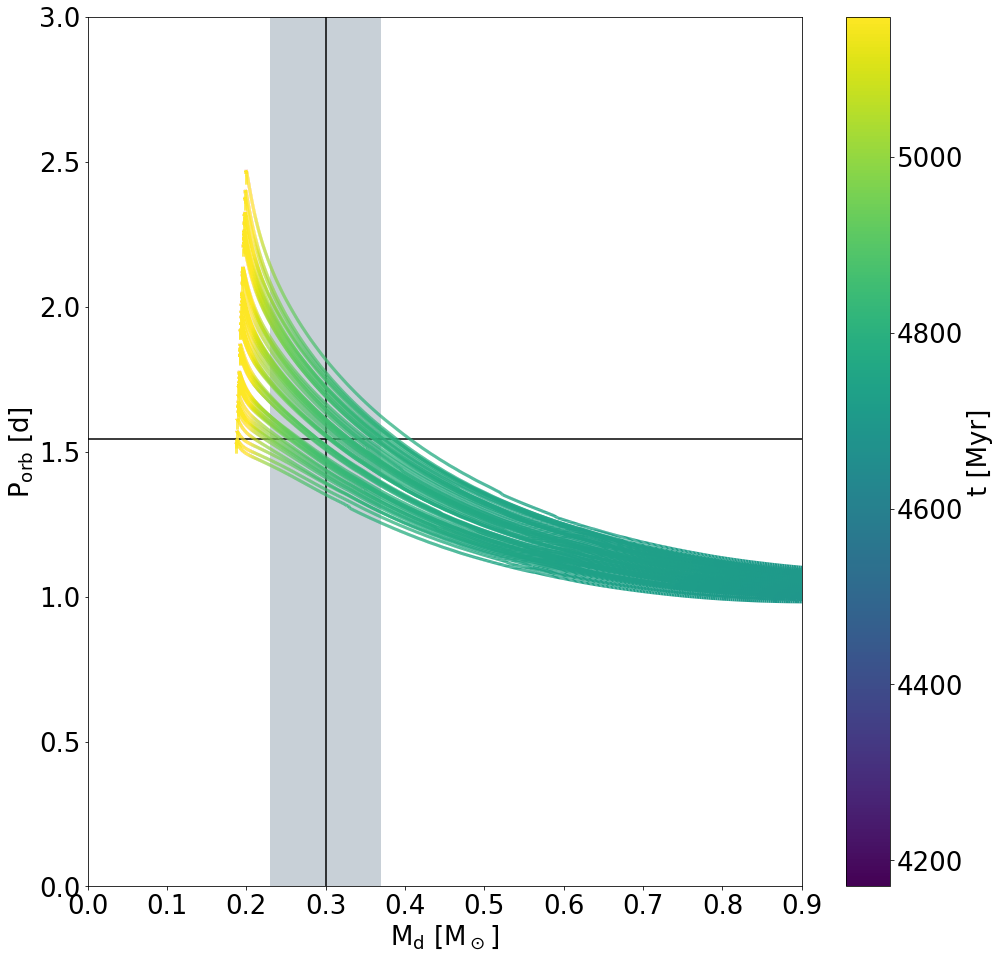}
       \end{subfigure}
       \hfill
       \begin{subfigure}{\columnwidth}
           \centering
           \includegraphics[width=\textwidth]{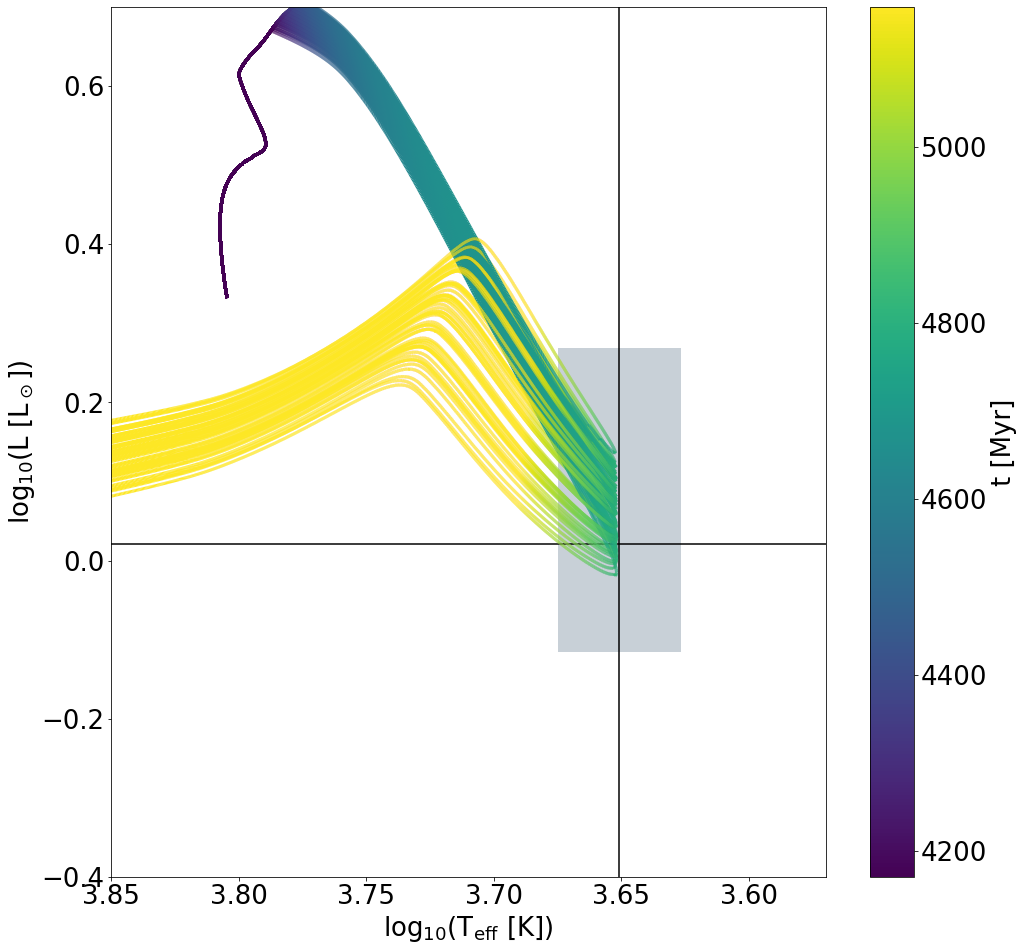}
       \end{subfigure}
       \newline
       \begin{subfigure}{\columnwidth}
           \centering
           \includegraphics[width=\textwidth]{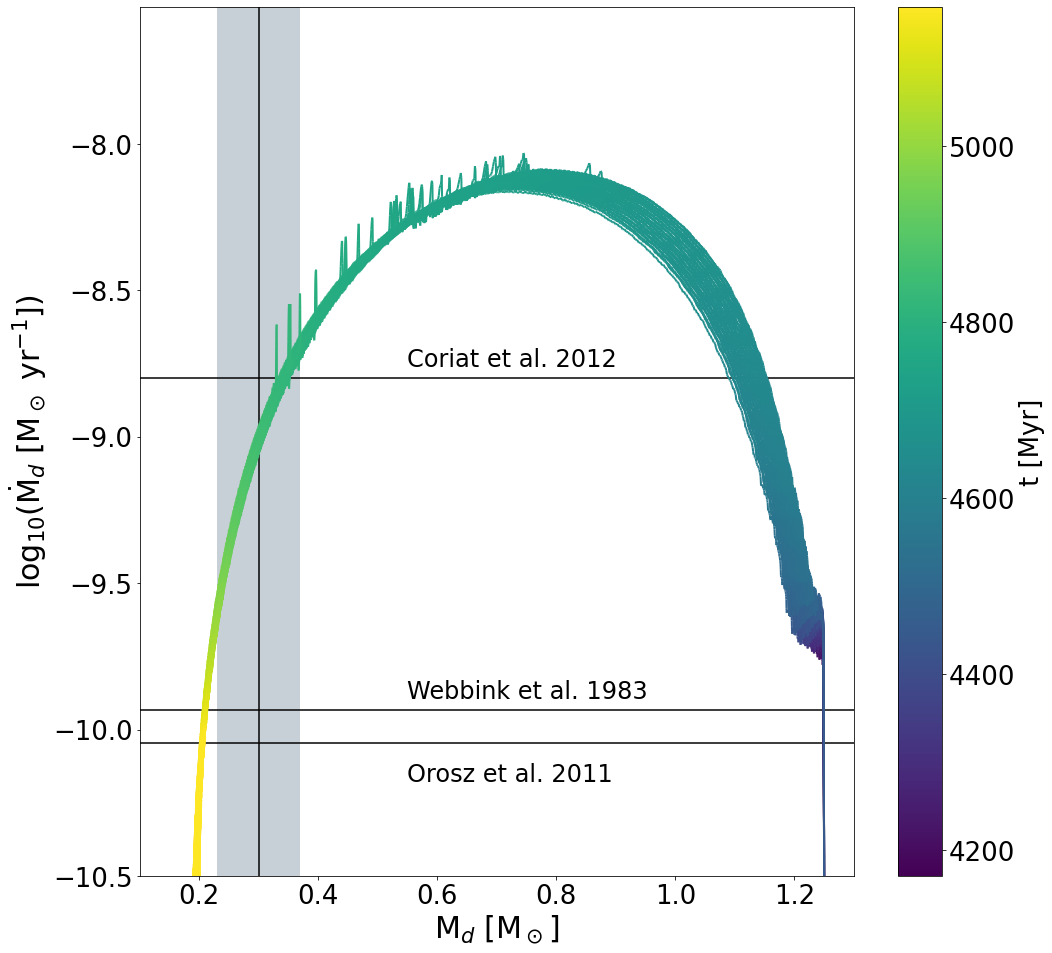}
       \end{subfigure}
       \hfill
       \begin{subfigure}{\columnwidth}
           \centering
           \includegraphics[width=\textwidth]{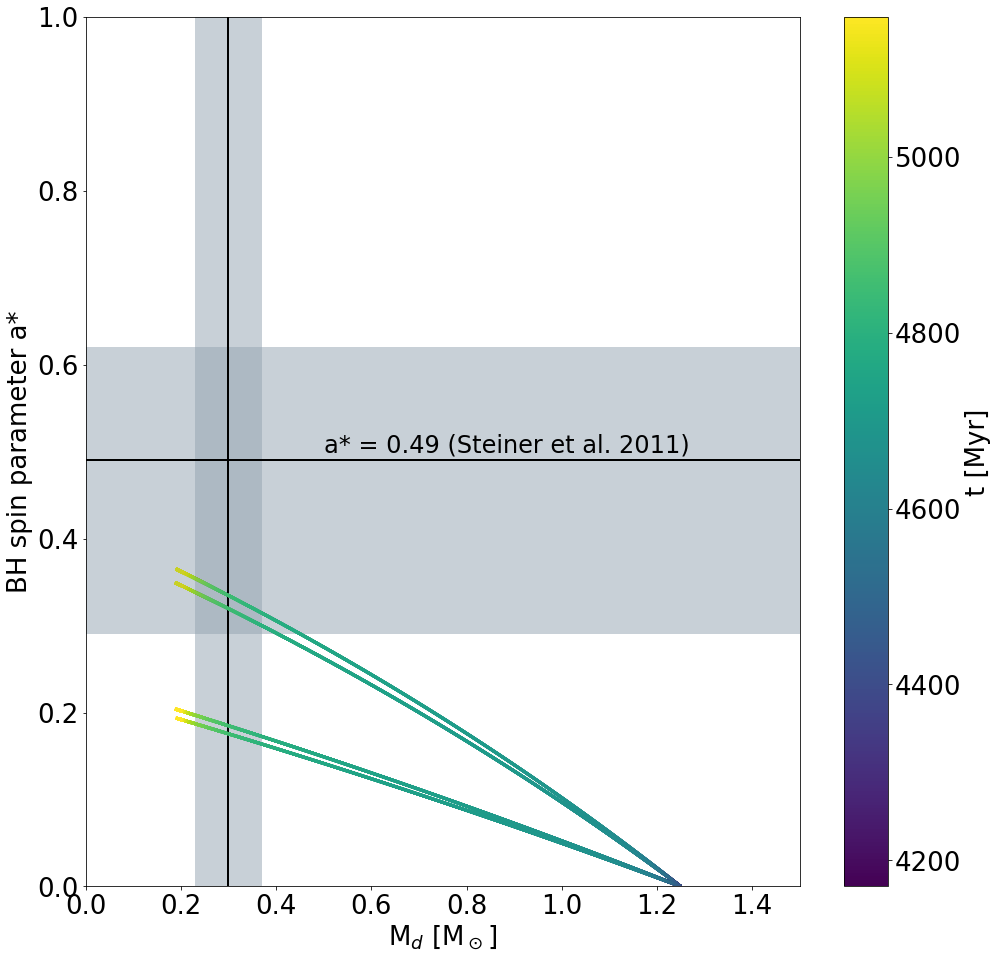}
       \end{subfigure}
       \caption{Same quantities as in Figure 5, but for the evolution of models computed using the CARB MB prescription.}
       \label{fig:CARB}
\end{figure*}
Another possibility for our models to reach this high value could be to change the MB prescription utilized. In the calculations presented in Section~\ref{sec:results}, we assumed the prescription derived by \cite{Verbunt81} and \cite{Rappaport83} based on the empirical Skumanich law \citep{1972ApJ...171..565S}. This prescription (MB0) was calibrated for main sequence stars similar to the Sun. However, it has been noted to face various issues when modeling the evolution of low-mass X-ray binaries with accreting neutron stars \citep{2003ApJ...597.1036P,2014A&A...571A..45I,2015ApJ...809...99S}, including the underestimation of mass loss rates compared to observed estimates for systems with a neutron star accretor \citep{Podsiadlowski2002,2003ApJ...597.1036P,2019ApJ...886L..31V}. This suggests that donor stars other than the Sun could exhibit increased wind-driven mass loss and magnetic fields whose strengths do not scale  with the rotational velocity in a straightforward way \citep{1968MNRAS.138..359M,10.1093/mnras/226.1.57,1988ApJ...333..236K}.

With this motivation, \cite{2019MNRAS.483.5595V} proposed three new MB prescriptions:  convection-boosted (MB2), intermediate (MB3), and wind-boosted (MB4). These prescriptions incorporate dependencies of magnetic field strength on wind mass-loss rate and the structure of the convective envelope. Later that same year, \cite{2019ApJ...886L..31V} presented a new MB prescription: CARB. This prescription also includes the effect of stellar rotation on the Alfvén radius. According to \cite{2019ApJ...886L..31V} and \cite{Echeveste2024}, the prescriptions that best represent LMXBs with a neutron star accretor are MB3 (based on orbital period, mass transfer rate, and mass ratio; although it fails to reproduce the effective temperature of Sco X-1) and CARB, which is able to more accurately describe the observed LMXB population and reproduce the observed mass transfer rates. This could also potentially be the case if the accretor were a BH.

In the context of XTE J1550-564, using a stronger MB prescription such as MB3 or CARB could  not only help us achieve higher mass transfer rates that could match the observational estimate data available for this quantity, but also could potentially allow the BH to accrete more matter and, thus, to increase  its spin parameter even further. With this in mind, a new search for progenitors of XTE~J1550-564 was necessary, since the evolution in binaries using different MBs can lead to a completely different orbital evolution of the system and, consequently, to different possible progenitors.

On the one hand, we explored the MB3 prescription, which, in contrast to the MB0, takes into account the effect of the convective envelope of the donor and its mass loss via winds. On the other hand, we considered the CARB prescription, which also includes the effect of the stellar rotation on the Alfvén radius. Both of them lead to stronger mass transfer rates than MB0 \citep{2021ApJ...909..174D,2024ApJ...971...54D,Echeveste2024}.

To design a new grid of models for each MB prescription, we first needed to restrain the initial parameter space. We carried out a wide exploration of models to delimit the ones that approach the observed features of XTE~J1550-564. The parameters that mostly affect the whole system's evolution are the initial orbital period and initial donor mass. Changing them directly impacts the system's orbital period by the time it reaches inside the donor's mass error margin or the position in the HR diagram, respectively. Once we know the values for the initial donor's mass and orbital period, we can design our grid to also explore the other two initial parameters: the initial BH mass and the $\beta$ parameter. For the case of MB3, we computed models with an initial donor mass of $0.96$~M$_\odot$ and initial orbital periods between 9.5 and 10.4~d. On the other hand, for the models employing the CARB prescription, the initial donor mass was set at $1.25$~M$_\odot$, with initial orbital periods ranging from 1.09 to 1.19~d. In both cases, we varied the initial BH mass between 8.5 and 9.0~M$_\odot$, and adopted values of the $\beta$ parameter equal to 0.5 (intermediate non-conservative case) and 1.0 (conservative case). Our results are summarized in Figures \ref{fig:MB3} and \ref{fig:CARB} for the models using the MB3 and CARB MB prescriptions, respectively.

In the case of MB3, the selected models reach the region of observed parameters at an advanced age, approximately 13.8~Gyr. These models begin with large orbital periods, as the strength of the MB efficiently drives the system toward shorter periods ($\sim$~1.5~d), even when the mass transfer episode has not begun. For all models, there is only a single, brief mass transfer episode lasting about 120~Myr. During this phase, both conservative and non-conservative models exhibit mass accretion rates that are narrowly limited by the critical Eddington mass accretion rate ($\approx 4\times10^{-7}$~M$_\odot$~yr$^{-1}$). When the donor star reaches the observed range of parameters, the mass loss rate drops significantly; however, it remains roughly one order of magnitude above the estimate from \citetalias{Coriat2012}.

For the models computed with the CARB prescription, the observed fundamental parameters of the system are reached at an age of approximately 4.9~Gyr. In this case, the binary components tend to move apart, so all models begin with initial orbital periods shorter than the currently observed one. The mass transfer episode is much longer than in the previous case, lasting about 1~Gyr. By the time the models reach the observed parameter region, the mass transfer rate aligns well with the observational estimate by \citetalias{Coriat2012} and the accretion rate stays below the Eddington limit.
\subsection{Considering whether an initially non-rotating BH explain the observed value of the BH spin parameter}
The primary objective of this study was to address this question and the findings in the MB0 context are shown in Figure \ref{fig:spin}. Model ``E'' reaches a value of $a^*$ just below the error bar of observations, while models with non-conservative mass transfer do not even reach half of the estimation. With the available observational data of this object, we cannot discard the occurrence of conservative mass transfer (noting that one argument in favor of this is the sub-Eddington mass transfer rate for this object). In addition, with the same reasoning, we cannot discard non-conservative mass transfer as the actual phenomenon occurring in XTE~J1550-564. These systems are renowned for experiencing mass loss through various mechanisms, including the ejection of relativistic jets, outflows from the accretion disk, and other processes \citep{Miller2015}. Notably, the XTE~J1550-564 system has a history of recurrent outburst events, during which the presence of relativistic jets has been inferred \citep{2001ApJ...554...43C,2001AIPC..587..126C}. Also, evidence of ionized disk winds has been found for this object by \cite{2020ApJ...892...47C}.

The study by \citetalias{Fragos2015} was a pioneering effort in exploring the origin of the BH spin parameter, focusing on the conservative mass transfer scenario and comparing their models to 9 BHXBs with $a^*$ measured via the continuum fitting technique. In this work, we focus our analysis on one of these systems, XTE~J1550-564, under more general assumptions and by performing detailed evolutionary models that represent all the observed parameters of this system simultaneously. Given that the evolution of $a^*$ is highly sensitive to the $\beta$ parameter, we consider it essential to explore this quantity as a free parameter. On the other hand, XTE~J1550-564 is particularly well-suited for our study, as its $a^*$ has been consistently measured using both techniques \citep{Steiner2011,2013MNRAS.431..405R}, resulting in a moderate value that makes it more favorable to address, considering an initially non-rotating BH.

Among our best-fitting models, one (model "E") closely aligns with the results of \citetalias{Fragos2015}, as it represents a conservative mass transfer case. However, even in this scenario, the predicted spin value of $a^* = 0.49$ remains inconsistent with all the models we computed using the MB0 prescription. More broadly, our analysis takes the non-conservative mass transfer as the most likely evolutionary pathway for XTE~J1550-564.\\
As for the exploration made using different MB prescriptions, the situation remains for the case of MB3. Even when the mass accretion rate exceeds the Eddington limit for both the conservative and non-conservative regimes, the entire event happens quickly enough to allow the BH accrete sufficient mass. Consequently, the BH cannot accelerate enough to reach the spin parameter. The evolution of the BH spin parameter for these models can be seen in Figure 5. As for the models computed using the CARB prescription, the amount of mass accreted by the BH allows the BH spin parameter to marginally reach within the error of the observation only in the conservative case (Figure 6).

Although MB prescriptions such as CARB seem to provide better agreement with observations in systems where the accretor is a neutron star, this does not appear to hold for BH LMXBs. In these systems, lower-efficiency MB prescriptions such as the Skumanich-based law tend to be more consistent with their global properties \citep{2024ApJ...971...54D}. Thus, although the CARB prescription appears promising for marginally reproducing the observed spin, its applicability to the XTE J1550-564 system remains uncertain.\\

These findings suggest that either some key assumptions need revision or certain physical processes are missing from our models. In particular, the hypothesis of an initially non-rotating BH could warrant reconsideration.
\section{Conclusions}\label{sec:conclusions}

In our previous study, we conducted a comprehensive analysis of the evolution of the donor star in the V404~Cyg system, a known BHXB. A significant finding was the system's inability to reach its estimated dimensionless BH spin parameter, $a^*$. Given this outcome, while the $a^*$ value was high and uncertain, we conjectured that a more confident and lower value of $a^*$ could be achieved using the same assumptions for another BHXB system. To test this hypothesis, we turned our attention to the system XTE~J1550-564, characterized by a moderate BH spin value of $a^* = 0.49$, measured with the agreement of both existing techniques for estimating this parameter.

With this aim, we computed the evolution of a few hundred systems and selected six that show a  very good agreement with the observational data available for the system. These models reproduce the current orbital period, donor and accretor masses, effective temperature, and luminosity of the donor star. The computed mass transfer rates are consistent with theoretical expectations \citep{Webbink83,King88}, but remain below the higher observational estimate based on X-ray luminosity \citepalias{Coriat2012}.

Assuming the MB0 prescription, based on the Skumanich law \citep{Verbunt81,Rappaport83}, none of our models can spin up the BH to the observed value for XTE~J1550-564. This constitutes that the observed spin of the BH in XTE~J1550-564 cannot be explained via accretion alone under standard circumstances.

These results differ if we consider a stronger MB prescription, which largely impacts the whole system evolution. To this end, we performed additional calculations using two alternative MB prescriptions: MB3 \citep{2019MNRAS.483.5595V} and CARB \citep{2019ApJ...886L..31V}. Models computed with MB3 result in extremely high mass transfer rates over short timescales, which prevent the BH from accreting enough material to increase its spin significantly. On the other hand, models that use the CARB prescription end upleading to longer mass transfer phases and higher spin gains. For this last case (the fully conservative case), the resulting spin values marginally reach the lower bound of the observed error range, although other authors \citep{2024ApJ...971...54D} do not tend to favor this prescription for BHXBs.

In any case, by not reaching the BH spin estimation (or only marginally reaching its lower bound), these findings seem to echo previous results for other systems (e.g., V404~Cyg), reinforcing the idea that the spin evolution of BH in X-ray binaries remains an open problem. The question of whether the inability to reproduce observed spins under standard assumptions is a general feature of BHXBs remains open. We intend to test this notion in a future work.

\begin{acknowledgements}
      The authors thank the anonymous referee for their comments, which were of great help in improving the original version of this work.
\end{acknowledgements}

\bibliographystyle{aa}
\bibliography{biblio}

\begin{thebibliography}{57}
\expandafter\ifx\csname natexlab\endcsname\relax\def\natexlab#1{#1}\fi

\bibitem[{Bahramian \& Degenaar(2022)}]{Bahramian2022}
Bahramian, A. \& Degenaar, N. 2022, Low-Mass X-ray Binaries (Singapore: Springer Nature Singapore), 1--62

\bibitem[{{Bardeen}(1970)}]{Bardeen70}
{Bardeen}, J.~M. 1970, \nat, 226, 64

\bibitem[{{Bartolomeo Koninckx} {et~al.}(2023){Bartolomeo Koninckx}, {De Vito}, \& {Benvenuto}}]{Bartolomeo2023}
{Bartolomeo Koninckx}, L., {De Vito}, M.~A., \& {Benvenuto}, O.~G. 2023, \aap, 674, A97

\bibitem[{{Benvenuto} \& {De Vito}(2003)}]{Benvenuto2003}
{Benvenuto}, O.~G. \& {De Vito}, M.~A. 2003, \mnras, 342, 50

\bibitem[{{Benvenuto} {et~al.}(2012){Benvenuto}, {De Vito}, \& {Horvath}}]{BDVHa}
{Benvenuto}, O.~G., {De Vito}, M.~A., \& {Horvath}, J.~E. 2012, \apjl, 753, L33

\bibitem[{{Blum} {et~al.}(2009){Blum}, {Miller}, {Fabian}, {Miller}, {Homan}, {van der Klis}, {Cackett}, \& {Reis}}]{Blum2009}
{Blum}, J.~L., {Miller}, J.~M., {Fabian}, A.~C., {et~al.} 2009, \apj, 706, 60

\bibitem[{{B{\"u}ning} \& {Ritter}(2004)}]{2004A&A...423..281B}
{B{\"u}ning}, A. \& {Ritter}, H. 2004, \aap, 423, 281

\bibitem[{{Carroll} \& {Ostlie}(2006)}]{2006ima..book.....C}
{Carroll}, B.~W. \& {Ostlie}, D.~A. 2006, {An introduction to modern astrophysics and cosmology}

\bibitem[{{Connors} {et~al.}(2020){Connors}, {Garc{\'\i}a}, {Dauser}, {Grinberg}, {Steiner}, {Sridhar}, {Wilms}, {Tomsick}, {Harrison}, \& {Licklederer}}]{2020ApJ...892...47C}
{Connors}, R. M.~T., {Garc{\'\i}a}, J.~A., {Dauser}, T., {et~al.} 2020, \apj, 892, 47

\bibitem[{{Corbel} {et~al.}(2001{\natexlab{a}}){Corbel}, {Kaaret}, {Jain}, {Bailyn}, {Fender}, {Tomsick}, {Kalemci}, {McIntyre}, {Campbell-Wilson}, {Miller}, \& {McCollough}}]{2001AIPC..587..126C}
{Corbel}, S., {Kaaret}, P., {Jain}, R.~K., {et~al.} 2001{\natexlab{a}}, in American Institute of Physics Conference Series, Vol. 587, Gamma 2001: Gamma-Ray Astrophysics, ed. S.~{Ritz}, N.~{Gehrels}, \& C.~R. {Shrader}, 126--130

\bibitem[{{Corbel} {et~al.}(2001{\natexlab{b}}){Corbel}, {Kaaret}, {Jain}, {Bailyn}, {Fender}, {Tomsick}, {Kalemci}, {McIntyre}, {Campbell-Wilson}, {Miller}, \& {McCollough}}]{2001ApJ...554...43C}
{Corbel}, S., {Kaaret}, P., {Jain}, R.~K., {et~al.} 2001{\natexlab{b}}, \apj, 554, 43

\bibitem[{{Coriat} {et~al.}(2012){Coriat}, {Fender}, \& {Dubus}}]{Coriat2012}
{Coriat}, M., {Fender}, R.~P., \& {Dubus}, G. 2012, \mnras, 424, 1991

\bibitem[{Cowley(1992)}]{Cowley92}
Cowley, A.~P. 1992, ARA\&A, 30, 287

\bibitem[{{De Vito} \& {Benvenuto}(2012)}]{DeVito2012}
{De Vito}, M.~A. \& {Benvenuto}, O.~G. 2012, \mnras, 421, 2206

\bibitem[{{Deng} \& {Li}(2024)}]{2024ApJ...971...54D}
{Deng}, Z.-L. \& {Li}, X.-D. 2024, \apj, 971, 54

\bibitem[{{Deng} {et~al.}(2021){Deng}, {Li}, {Gao}, \& {Shao}}]{2021ApJ...909..174D}
{Deng}, Z.-L., {Li}, X.-D., {Gao}, Z.-F., \& {Shao}, Y. 2021, \apj, 909, 174

\bibitem[{{Echeveste} {et~al.}(2024){Echeveste}, {Novarino}, {Benvenuto}, \& {De Vito}}]{Echeveste2024}
{Echeveste}, M., {Novarino}, M.~L., {Benvenuto}, O.~G., \& {De Vito}, M.~A. 2024, \mnras, 530, 4277

\bibitem[{{Eggleton}(1983)}]{Eggleton83}
{Eggleton}, P.~P. 1983, \apj, 268, 368

\bibitem[{{Fabian} {et~al.}(1989){Fabian}, {Rees}, {Stella}, \& {White}}]{Fabian89}
{Fabian}, A.~C., {Rees}, M.~J., {Stella}, L., \& {White}, N.~E. 1989, \mnras, 238, 729

\bibitem[{{Fragos} \& {McClintock}(2015)}]{Fragos2015}
{Fragos}, T. \& {McClintock}, J.~E. 2015, \apj, 800, 17, hereafter referred to as F+15

\bibitem[{{Gou} {et~al.}(2011){Gou}, {McClintock}, {Reid}, {Orosz}, {Steiner}, {Narayan}, {Xiang}, {Remillard}, {Arnaud}, \& {Davis}}]{Gou2011}
{Gou}, L., {McClintock}, J.~E., {Reid}, M.~J., {et~al.} 2011, \apj, 742, 85

\bibitem[{{Hameury} {et~al.}(1993){Hameury}, {King}, {Lasota}, \& {Raison}}]{1993A&A...277...81H}
{Hameury}, J.~M., {King}, A.~R., {Lasota}, J.~P., \& {Raison}, F. 1993, \aap, 277, 81

\bibitem[{{Istrate} {et~al.}(2014){Istrate}, {Tauris}, \& {Langer}}]{2014A&A...571A..45I}
{Istrate}, A.~G., {Tauris}, T.~M., \& {Langer}, N. 2014, \aap, 571, A45

\bibitem[{{Kawaler}(1988)}]{1988ApJ...333..236K}
{Kawaler}, S.~D. 1988, \apj, 333, 236

\bibitem[{{King}(1988)}]{King88}
{King}, A.~R. 1988, \qjras, 29, 1

\bibitem[{{King} \& {Kolb}(1999)}]{King99}
{King}, A.~R. \& {Kolb}, U. 1999, \mnras, 305, 654

\bibitem[{{Kippenhahn} \& {Weigert}(1967)}]{1967ZA.....65..251K}
{Kippenhahn}, R. \& {Weigert}, A. 1967, \zap, 65, 251

\bibitem[{Kretschmar {et~al.}(2019)Kretschmar, Fürst, Sidoli, Bozzo, Alfonso-Garzón, Bodaghee, Chaty, Chernyakova, Ferrigno, Manousakis, Negueruela, Postnov, Paizis, Reig, Rodes-Roca, Tsygankov, Bird, {Bissinger né Kühnel}, Blay, Caballero, Coe, Domingo, Doroshenko, Ducci, Falanga, Grebenev, Grinberg, Hemphill, Kreykenbohm, {Kreykenbohm né Fritz}, Li, Lutovinov, Martínez-Núñez, Mas-Hesse, Masetti, McBride, Neronov, Pottschmidt, Rodriguez, Romano, Rothschild, Santangelo, Sguera, Staubert, Tomsick, Torrejón, Torres, Walter, Wilms, Wilson-Hodge, \& Zhang}]{KRETSCHMAR2019}
Kretschmar, P., Fürst, F., Sidoli, L., {et~al.} 2019, New Astronomy Reviews, 86, 101546

\bibitem[{{Landau} \& {Lifshitz}(1971)}]{Landau71}
{Landau}, L.~D. \& {Lifshitz}, E.~M. 1971, {The classical theory of fields} (Pergamon Press)

\bibitem[{{Laor}(1991)}]{Laor91}
{Laor}, A. 1991, \apj, 376, 90

\bibitem[{{McClintock} {et~al.}(2006){McClintock}, {Shafee}, {Narayan}, {Remillard}, {Davis}, \& {Li}}]{McClintock2006}
{McClintock}, J.~E., {Shafee}, R., {Narayan}, R., {et~al.} 2006, \apj, 652, 518

\bibitem[{{Mestel}(1968)}]{1968MNRAS.138..359M}
{Mestel}, L. 1968, \mnras, 138, 359

\bibitem[{Mestel \& Spruit(1987)}]{10.1093/mnras/226.1.57}
Mestel, L. \& Spruit, H.~C. 1987, Monthly Notices of the Royal Astronomical Society, 226, 57

\bibitem[{{Miller} {et~al.}(2015){Miller}, {Fabian}, {Kaastra}, {Kallman}, {King}, {Proga}, {Raymond}, \& {Reynolds}}]{Miller2015}
{Miller}, J.~M., {Fabian}, A.~C., {Kaastra}, J., {et~al.} 2015, \apj, 814, 87

\bibitem[{{Orosz} {et~al.}(1998){Orosz}, {Bailyn}, \& {Jain}}]{Orosz98}
{Orosz}, J., {Bailyn}, C., \& {Jain}, R. 1998, \iaucirc, 7009, 1

\bibitem[{{Orosz} {et~al.}(2002){Orosz}, {Groot}, {van der Klis}, {McClintock}, {Garcia}, {Zhao}, {Jain}, {Bailyn}, \& {Remillard}}]{Orosz2002}
{Orosz}, J.~A., {Groot}, P.~J., {van der Klis}, M., {et~al.} 2002, \apj, 568, 845

\bibitem[{Orosz {et~al.}(2011)Orosz, Steiner, McClintock, Torres, Remillard, Bailyn, \& Miller}]{Orosz2011}
Orosz, J.~A., Steiner, J.~F., McClintock, J.~E., {et~al.} 2011, \apj, 730, 75, hereafter referred to as O11

\bibitem[{{Pfahl} {et~al.}(2003){Pfahl}, {Rappaport}, \& {Podsiadlowski}}]{2003ApJ...597.1036P}
{Pfahl}, E., {Rappaport}, S., \& {Podsiadlowski}, P. 2003, \apj, 597, 1036

\bibitem[{{Podsiadlowski} {et~al.}(2003){Podsiadlowski}, {Rappaport}, \& {Han}}]{Podsiadlowski2003}
{Podsiadlowski}, P., {Rappaport}, S., \& {Han}, Z. 2003, \mnras, 341, 385

\bibitem[{{Podsiadlowski} {et~al.}(2002){Podsiadlowski}, {Rappaport}, \& {Pfahl}}]{Podsiadlowski2002}
{Podsiadlowski}, P., {Rappaport}, S., \& {Pfahl}, E.~D. 2002, \apj, 565, 1107

\bibitem[{{Rappaport} {et~al.}(1982){Rappaport}, {Joss}, \& {Webbink}}]{Rappaport82}
{Rappaport}, S., {Joss}, P.~C., \& {Webbink}, R.~F. 1982, \apj, 254, 616

\bibitem[{{Rappaport} {et~al.}(1983){Rappaport}, {Verbunt}, \& {Joss}}]{Rappaport83}
{Rappaport}, S., {Verbunt}, F., \& {Joss}, P.~C. 1983, \apj, 275, 713

\bibitem[{{R{\'e}ville} {et~al.}(2015){R{\'e}ville}, {Brun}, {Matt}, {Strugarek}, \& {Pinto}}]{2015ApJ...798..116R}
{R{\'e}ville}, V., {Brun}, A.~S., {Matt}, S.~P., {Strugarek}, A., \& {Pinto}, R.~F. 2015, \apj, 798, 116

\bibitem[{{Reynolds}(2021)}]{Reynolds2021}
{Reynolds}, C.~S. 2021, \araa, 59, 117

\bibitem[{{Ritter}(2008)}]{2008NewAR..51..869R}
{Ritter}, H. 2008, \nar, 51, 869

\bibitem[{{Russell} {et~al.}(2013){Russell}, {Gallo}, \& {Fender}}]{2013MNRAS.431..405R}
{Russell}, D.~M., {Gallo}, E., \& {Fender}, R.~P. 2013, \mnras, 431, 405

\bibitem[{{Shao} \& {Li}(2015)}]{2015ApJ...809...99S}
{Shao}, Y. \& {Li}, X.-D. 2015, \apj, 809, 99

\bibitem[{{Skumanich}(1972)}]{1972ApJ...171..565S}
{Skumanich}, A. 1972, \apj, 171, 565

\bibitem[{{Smith}(1998)}]{Smith98}
{Smith}, D.~A. 1998, \iaucirc, 7008, 1

\bibitem[{{Steiner} {et~al.}(2011){Steiner}, {Reis}, {McClintock}, {Narayan}, {Remillard}, {Orosz}, {Gou}, {Fabian}, \& {Torres}}]{Steiner2011}
{Steiner}, J.~F., {Reis}, R.~C., {McClintock}, J.~E., {et~al.} 2011, \mnras, 416, 941

\bibitem[{{Van} \& {Ivanova}(2019)}]{2019ApJ...886L..31V}
{Van}, K.~X. \& {Ivanova}, N. 2019, \apjl, 886, L31

\bibitem[{{Van} {et~al.}(2019){Van}, {Ivanova}, \& {Heinke}}]{2019MNRAS.483.5595V}
{Van}, K.~X., {Ivanova}, N., \& {Heinke}, C.~O. 2019, \mnras, 483, 5595

\bibitem[{Verbunt(1993)}]{Verbunt93}
Verbunt, F. 1993, ARA\&A, 31, 93

\bibitem[{{Verbunt} \& {Zwaan}(1981)}]{Verbunt81}
{Verbunt}, F. \& {Zwaan}, C. 1981, \aap, 100, L7

\bibitem[{{Walton} {et~al.}(2017){Walton}, {Mooley}, {King}, {Tomsick}, {Miller}, {Dauser}, {Garc{\'\i}a}, {Bachetti}, {Brightman}, {Fabian}, {Forster}, {F{\"u}rst}, {Gandhi}, {Grefenstette}, {Harrison}, {Madsen}, {Meier}, {Middleton}, {Natalucci}, {Rahoui}, {Rana}, \& {Stern}}]{Walton2017}
{Walton}, D.~J., {Mooley}, K., {King}, A.~L., {et~al.} 2017, \apj, 839, 110

\bibitem[{{Webbink} {et~al.}(1983){Webbink}, {Rappaport}, \& {Savonije}}]{Webbink83}
{Webbink}, R.~F., {Rappaport}, S., \& {Savonije}, G.~J. 1983, \apj, 270, 678

\bibitem[{{Zhang} {et~al.}(1997){Zhang}, {Cui}, \& {Chen}}]{Zhang97}
{Zhang}, S.~N., {Cui}, W., \& {Chen}, W. 1997, \apjl, 482, L155

\end{thebibliography}

\end{document}